\documentclass[a4paper,11pt]{article}
\pdfoutput=1 

\usepackage{jcappub} 

\usepackage[T1]{fontenc} 
\usepackage{empheq}
\usepackage{listings}
\usepackage{float}
\usepackage{placeins}
\usepackage{graphicx} 
\usepackage{tablefootnote}
\usepackage{array}
\usepackage{booktabs}
\usepackage{siunitx}
\usepackage{makecell}
\usepackage{hyperref}
\usepackage{cleveref}
\usepackage{multirow}
\usepackage{xparse}

\DeclareSIUnit{\pc}{pc}
\DeclareSIUnit{\kpc}{\kilo\pc}
\DeclareSIUnit{\Mpc}{\mega\pc}
\DeclareSIUnit{\hHubble}{\text{\ensuremath{h}}}

\newcommand{\vecx}{\mathbf{x}}
\newcommand{\vecr}{\mathbf{r}}

\newcommand{\dK}{\delta^{\rm K}}

\newcommand{\hatr}{\widehat{\mathbf{r}}}

\newcommand{\hatn}{\widehat{\mathbf{n}}}

\newcommand{\PP}{\mathcal{P}}
\newcommand{\tj}[6]{\begin{pmatrix} {#1} & {#2} & {#3}\\ {#4} & {#5} & {#6}\end{pmatrix}}

\makeatletter
\let\jnl@style=\relax
\makeatother

\title{\boldmath A Two-Dimensional Test of Cosmological Parity Violation with the 3-Point Correlation Function}

\author[1]{William Ortolá Leonard}
\author[2]{\& Zachary Slepian}

\affiliation[1]{Department of Physics, University of Florida,\\2001 Museum Rd., Gainesville, FL 32611, USA}
\affiliation[2]{Department of Astronomy, University of Florida,\\211 Bryant Space Science Center, Gainesville, FL 32611, USA}

\emailAdd{wortola@ufl.edu, zslepian@ufl.edu}

\usepackage{empheq}
\usepackage{listings}
\usepackage{hyperref}
\usepackage{tablefootnote}
\usepackage{graphicx} 

\abstract{Searches for parity violation (PV) in the Universe's large-scale structure have recently been carried out using parity-odd modes of the galaxy 4-Point Correlation Function (4PCF). Here, we introduce a new estimator for a 3-Point Correlation Function (3PCF) that is sensitive to PV. The construction exploits the handedness of triangles formed on 2D spherical shells, where mirror-image configurations cannot be related by proper rotations within the surface. By dotting the oriented area vector with an externally-defined normal vector to the shell, we obtain sensitivity to PV. We present a discrete estimator for this statistic and also show how to correct for the survey geometry. This parity-odd 3PCF provides a complementary test of PV in galaxy surveys and a potential cross-check of PV 4PCF measurements.}

\begin{document}
\maketitle
\flushbottom

\section{Introduction} \label{sec:intro}
In three spatial dimensions (3D), the 4-Point Correlation Function (4PCF) is the lowest-order N-Point Correlation Function (NPCF) capable of probing parity violation (PV) in the Universe's large-scale structure (LSS) \cite{cahn_parity}\footnote{\cite{shiraishii} presented the first dedicated study of parity violation in the CMB trispectrum.}. Generically, four points generically form a tetrahedron, which the simplest 3D object that cannot be rotated into its mirror image. Under a parity transformation ($\mathbf{x} \rightarrow -\mathbf{x}$), the orientation of the tetrahedron is inverted, and this inversion cannot be undone by a rotation.\footnote{Therefore, the 4PCF is sensitive to intrinsic handedness in the underlying density field. We note that a parity transformation is obtained through a mirror reflection and a rotation. However, in LSS analyses the correlation functions are averaged over rotations. Therefore, we refer to parity transformations as mirror reflections.}

Parity-odd modes in the 4PCF have been measured recently in several datasets. Using the Sloan Digital Sky Survey (SDSS) Baryon Oscillation Spectroscopic Survey (BOSS)  constant stellar mass (CMASS) sample of luminous red galaxies (LRGs), \cite{hou_parity} found up to $7.1\sigma$ evidence, while \cite{phil_parity} reported $2.9\sigma$ evidence.\footnote{\cite{hou_parity} also studied  the lower-redshift LOWZ sample of BOSS, finding more modest evidence consistent with this sample's smaller volume. \cite{phil_parity} used coarser radial bins than the fiducial analysis of \cite{hou_parity}, possibly explaining the more modest evidence (see \cite{hou_parity} \S2.2 for further discussion). \cite{hou_parity} also explored the same radial binning as \cite{phil_parity} and found consistent results with the latter.} Subsequent work on BOSS CMASS by \cite{Krolewski_No_PV_det} found between 0 and 2.5$\sigma$ evidence for a parity-violating signal in the same dataset, using a new cross-correlation method (see their $\S4.3$ for further details) that is robust to miss-estimate of the covariance matrix and mismatch between data and mock catalogs. 

More recently, \cite{Slepian_Parity_Meas_DESI} used the Dark Energy Spectroscopic Instrument (DESI) Year~1 (Y1) Luminous Red Galaxy (LRG) sample and found detections up to $10\sigma$ in the auto-correlation analysis (see their Methods section for further details). They also used the cross-correlation method proposed by \cite{Krolewski_No_PV_det} and found no evidence of PV. Later, \cite{Hou_Parity_odd_2} used the same DESI LRG sample and found at most $4\sigma$ evidence of PV via the auto-correlation method with coarser radial bins than \cite{Slepian_Parity_Meas_DESI}. \cite{Hou_Parity_odd_2} also performed a cross-correlation analysis and proposed a new method sensitive to data-mock mismatch (see their $\S$III for further details); they term this method the corrected auto-correlation analysis. Another parity study on DESI Y1 data, \cite{Gao_Compressed_PV_Search}, applied a compressed 4-point statistical approach---parity-odd kurto spectra---to search for PV.\footnote{They analyzed BOSS data as well, finding between $-4.3$ and $0.9\sigma$.} They find significance of $0.8-3.1\sigma$ in the auto-correlation analysis and $0.5-2.8\sigma$ significance in the cross-correlation analysis. These intriguing results motivate further independent validation.

Related parity-odd 3-point statistics have also been discussed in the context of the Cosmic Microwave Background (CMB) and redshift-space galaxy clustering. In the CMB, the bispectrum---the Fourier-space version of the 3-Point Correlation Function (3PCF)---correlates three spherical-harmonic modes, $\langle a_{\ell_1 m_1}a_{\ell_2 m_2}a_{\ell_3 m_3}\rangle$ \cite{spergel_goldberg_bispec, Golberg_Spergel_CMB_II, Luo_bispectrum_CMB}. \cite{Kamionkowski_Odd_Bispectrum} pointed out that the usual restriction $\ell_1+\ell_2+\ell_3=\mathrm{even}$ follows from assuming parity invariance, and that configurations with $\ell_1+\ell_2+\ell_3=\mathrm{odd}$ offer an odd-parity CMB bispectrum. Since the CMB is observed as a 2D field on the sky, these odd-parity bispectra distinguish triangles of opposite handedness in harmonic space. \cite{Kamionkowski_Odd_Bispectrum} suggested that such signals could arise, in principle, through weak lensing by chiral gravitational waves or through cosmological birefringence, although the expected amplitudes are likely small. Nevertheless, \cite{Kamionkowski_Odd_Bispectrum} emphasized that odd-parity CMB bispectra are valuable as null tests of standard even-parity CMB bispectrum analyses.

In galaxy clustering, the situation is more subtle because observed data contain a preferred direction, the line-of-sight, leading to redshift-space distortions (RSD) \cite{kaiser_1987, hamilton_1998, Jackson_FOG}. \cite{Jeong_Schmidt_odd_parity_Bisp} showed that the imaginary part of the redshift-space galaxy bispectrum is parity odd and can be generated even in $\Lambda$CDM by galaxy velocities through RSD. They concluded that this signal is expected to be small, with signal-to-noise at most of order unity for realistic surveys. They also conclude that it provides a clean consistency test and could become more important in modified-gravity scenarios \cite{Hou_MG_review}. \cite{Jeong_Schmidt_odd_parity_Bisp} work shows that odd-parity 3-point statistics are well motivated, but they also highlight the need to separate intrinsic parity violation from RSD.

In contrast with the 3D case, within two spatial dimensions (2D) the minimal shape analogous to a tetrahedron is a triangle, captured by the 3PCF. A triplet of points in 2D forms a triangle, and a triangle in a plane cannot be rotated onto its mirror image within that same plane. Only by leaving the plane---i.e., using the third dimension---could one rotate the mirror image back onto the original triangle. Thus, in a strictly 2D analysis, the 3PCF is the lowest-order statistic capable of detecting PV signals.  

In this work, we present a 3PCF estimator that could be used to test for PV in the LSS by measuring the 3PCF on 2D spherical shells. This analysis is particularly interesting to apply to ground-based photometric surveys such as the Legacy Survey of Space and Time (LSST) \cite{Ivezi_2019_LSSTReview}, as it observes galaxies in spherical shells (photometric bins) around us. At the same time, it is also suitable to apply to ground-based spectroscopic surveys such as Dark Energy Spectroscopic Instrument (DESI) \cite{DESI:2016}; or space-based surveys such as the Euclid Satellite Mission \cite{Euclid_overview}, the Spectro-Photometer for the History of the Universe, Epoch of Reionization and Ices Explorer Mission (SPHEREx) \cite{spherex}, and the Nancy Grace Roman Space Telescope (ROMAN) \cite{roman_sn}.

This work is structured as follows. In \Cref{sec:3PCF_Estimator}, we introduce a new 3PCF decomposition and demonstrate its sensitivity to parity. We then present the discrete estimator in \Cref{sec:Discrete_Estimator} and how to correct for the survey geometry in \Cref{sec:Edge_Corrections}. Finally, in \Cref{sec:conclusion} we summarize our findings and discuss directions for future work.

\section{The Parity-Violating 3-Point Correlation Function}\label{sec:3PCF_Estimator}
The isotropic 3PCF, expressed in terms of the Legendre polynomials $\mathcal{L}$, can be written as \cite{se_3pt_alg, Kamalinejad_BAO_3PCF_DESI, se_rv, se_boss_3pcf, se_3pcf_rsd, se_3pcf_bao, Istvan_3PCF,pan_3pcf}\footnote{Algorithms to compute the 3PCF, such as \textsc{ENCORE} \cite{encore}, uses the isotropic basis functions instead, which for two arguments  are related to Legendre polynomials by a normalization and a phase \cite{cahn_iso}.}
\begin{align} \label{eq:even_3PCF_standard}
\zeta(\vecr_1,\vecr_2) = \sum_{\ell} \zeta_{\ell}(r_1,r_2)\; \mathcal{L}_{\ell}(\hatr_1\cdot\hatr_2), 
\end{align} 
where the angular dependence is entirely encoded in the dot product $\hatr_1\cdot\hatr_2$. As a result, this decomposition is insensitive to the orientation of the triangle and removes any parity-odd signal. Anisotropic 3PCF or bispectrum models \cite{se_aniso_3pcf, Sugiyama_Anisotropic_Bispectrum, Jeong_Schmidt_odd_parity_Bisp} may in principle contain terms that appear to mimic a parity-violating signal; however, these terms are tied to the line-of-sight dependence and must vanish once rotational invariance (isotropy) is enforced. The challenge is therefore to construct a 3PCF estimator that remains rotationally invariant while still being sensitive to parity violation. 

With this goal in mind, we introduce an odd-parity 3PCF expanded as
\begin{align}\label{eq:PV_3PCF_Expansion} 
\zeta(\hatn,\vecr_1,\vecr_2) = \sum_{\ell_n,\ell_1,\ell_2}\zeta_{\ell_n,\ell_1,\ell_2}(r_1,r_2) \PP_{\ell_n\ell_1\ell_2}(\hatn,\hatr_1,\hatr_2), 
\end{align} 
where $\hat{\mathbf n}$ denotes an external normal vector associated with the spherical shell. The coefficients $\zeta_{\ell_n,\ell_1,\ell_2}(r_1,r_2)$ are the radial multiplets obtained after integrating over the angular dependence of $(\hat{\mathbf n},\hat{\mathbf r}_1,\hat{\mathbf r}_2)$. In this sense, \cref{eq:PV_3PCF_Expansion} is the analog of the usual isotropic 3PCF expansion, but extended to include the external orientation needed to distinguish the handedness of the triangles. This expansion exploits the isotropic basis functions of \cite{cahn_iso}---defined in \cref{eq:Iso_function_def}.

The full 3PCF may be schematically decomposed into parity-even and parity-odd contributions,
\begin{equation}\label{eq:even_odd_decomposition}
    \zeta(\hat{\mathbf n},\mathbf r_1,\mathbf r_2)
    =
    \zeta_{\rm even}(\hat{\mathbf n},\mathbf r_1,\mathbf r_2)
    +
    \zeta_{\rm odd}(\hat{\mathbf n},\mathbf r_1,\mathbf r_2).
\end{equation}
The standard isotropic 3PCF of \cref{eq:even_3PCF_standard} is sensitive only to the parity-even angular dependence. In this work, we focus on the parity-odd component, which changes sign under parity transformation. 

\begin{figure}
    \centering
    \includegraphics[width=0.9\linewidth]{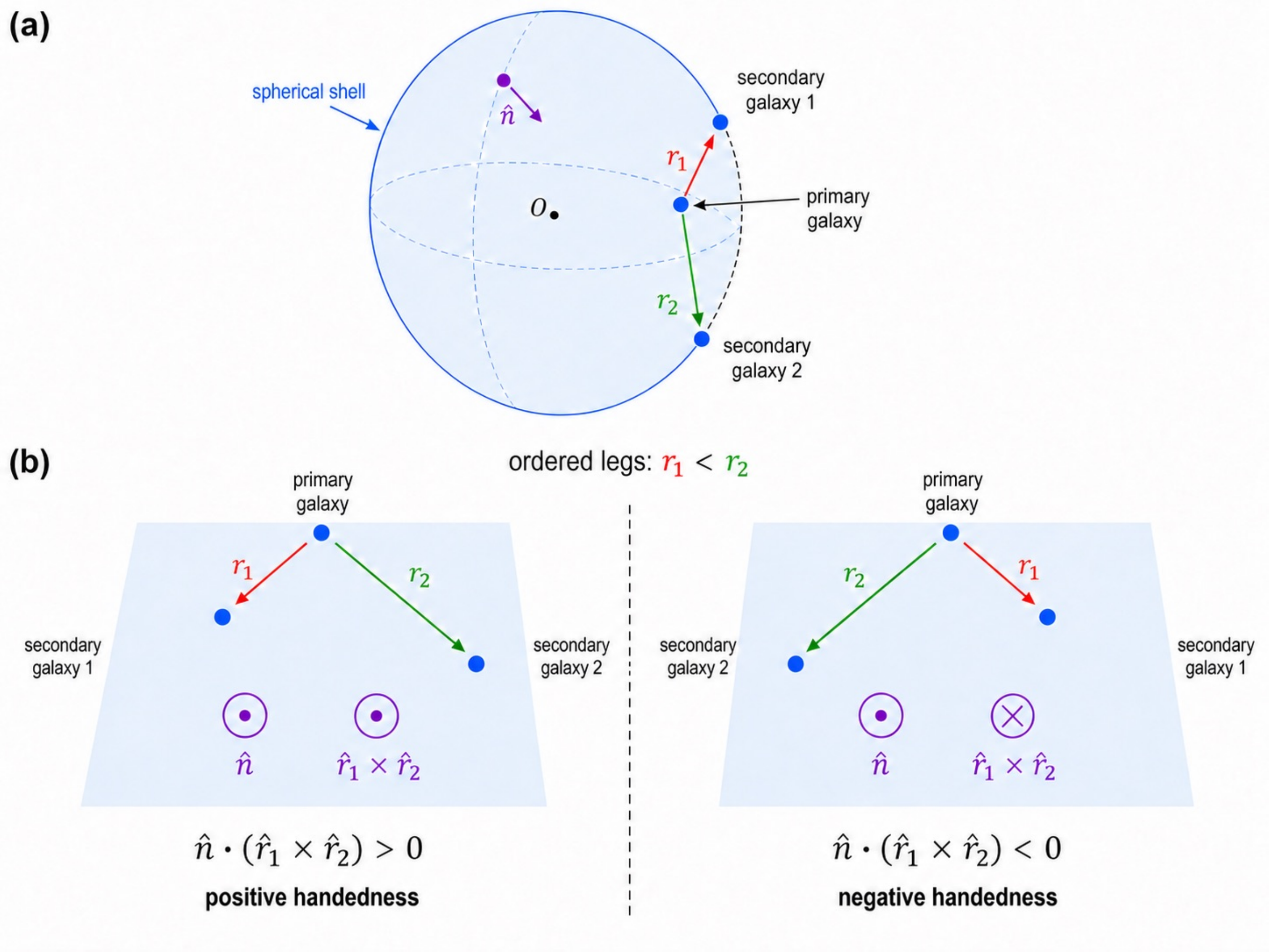}
    \caption{Visualization of the parity-odd 3PCF estimator. Panel (a) shows three galaxies located on a 2D spherical shell: a primary galaxy, and two secondary galaxies, which latter define the triangle legs $\mathbf r_1$ and $\mathbf r_2$, with $r_1<r_2$. $\hat{\mathbf n}$ denotes an externally-defined normal vector to the shell, shown here pointing inward toward the observer \textit{O} at the shell's center. Panel (b) illustrates the handedness test from the local 2D representation of the shell. The circled dot denotes a vector pointing pointing inside of the spherical shell, while the circled cross denotes a vector pointing outside of the spherical shell. Configurations with $\hat{\mathbf n}\cdot(\hat{\mathbf r}_1\times\hat{\mathbf r}_2)>0$ define positive handedness (panel (b), left), while mirror configurations with $\hat{\mathbf n}\cdot(\hat{\mathbf r}_1\times\hat{\mathbf r}_2)<0$ define negative handedness (panel (b), right).The placement of $\hat{\mathbf  n}$ in panel (b) is illustrative and is meant only to illustrate the sign of the scalar triple product.}
    \label{fig:PV_3PCF_Mehcanism}
\end{figure}

The parity-odd component is selected using the kernel
\begin{equation}\label{eq:pv_kernel}
f(\hat{\mathbf n},\hat{\mathbf r}_1,\hat{\mathbf r}_2)
\equiv
\hat{\mathbf n}\cdot(\hat{\mathbf r}_1\times \hat{\mathbf r}_2)
=
\mathcal{D}\;
\mathcal{P}_{111}(\hat{\mathbf n},\hat{\mathbf r}_1,\hat{\mathbf r}_2),
\end{equation}
where 
\begin{align} 
\mathcal{D} \equiv \frac{i \sqrt{2}\;(4\pi)^{3/2}}{3 V}, 
\end{align} 
and the second equality follows from Eq.~(A1) in \cite{hou_parity}. $\hat{\mathbf r}_1\times\hat{\mathbf r}_2$ is the normal vector of the triangle, which changes orientation under a parity transformation. By dotting the normal vector of the triangle with $\hatn$---which maintains its orientation under a parity transformation---$f$ captures the handedness of the triangle. This behavior is schematically described in \Cref{fig:PV_3PCF_Mehcanism}. 

We then define the parity-odd 3PCF in terms of the density fluctuation field, $\delta(\mathbf x)=\rho(\mathbf x)/\bar{\rho}-1$ (where $\rho(\vecx)$ is the density at $\vecx$ and $\bar{\rho}$ is the mean density), as
\begin{align}\label{eq:PV_3PCF_Def} 
\zeta(\hatn,\vecr_1,\vecr_2) \equiv \frac{f(\hatn, \hatr_1,\hatr_2)}{V} \int d^3\vecx \;\delta (\vecx) \delta(\vecx+\vecr_1)\delta(\vecx+\vecr_2), \end{align} 
where $V$ is the survey volume and we have dropped the subscript indicating this is the odd contribution. Then, \cref{eq:PV_3PCF_Def} is the definition of the parity-odd contribution appearing in \cref{eq:even_odd_decomposition}. 

The parity-odd 3PCF above, however, is a 3D estimator. The goal of this work is to construct a parity-odd 3PCF estimator restricted to a 2D spherical shell. Hence, we define a spherical tophat window function as
\begin{align}
    W_{R}(\vecx) \equiv H(|\vecx|-R_{\rm min})H(R_{\rm max}-|\vecx|),
\end{align}
where $H$ is the Heaviside function. In the ideal 2D shell limit $R_{\rm max}\rightarrow R_{\rm min}$; \textit{i.e.,} all galaxies lie perfectly on the surface of a sphere of radius $R$. However, in a real survey not all galaxies will lie exactly at the same distance $R$ from an observer and we must account for this fact. The parity-odd 3PCF estimator constrained to a 2D spherical shell then becomes
\begin{align}\label{eq:Shell_PV_3PCF}
&\zeta_{R}(\hatn,\vecr_1,\vecr_2) \equiv \frac{f(\hatn, \hatr_1,\hatr_2)}{V_R} \int d^3\vecx \;W_{R}(\vecx)\delta (\vecx) \nonumber \\
&\qquad \qquad \qquad \times W_{R}(\vecx+\vecr_1)\delta(\vecx+\vecr_1)\;W_{R}(\vecx+\vecr_2)\delta(\vecx+\vecr_2),   
\end{align}
where $V_{R}$ is the shell volume between $R_{\rm min}$ and $R_{\rm max}$.

To isolate the radial coefficients of \cref{eq:PV_3PCF_Expansion}, we substitute in \Cref{eq:Shell_PV_3PCF} and then exploit the isotropic basis functions' orthogonality to obtain
\begin{align}\label{eq:Odd3PCF_Continuous_Estimator} 
&\zeta_{\ell_n,\ell_1,\ell_2; R}(r_1,r_2) = \mathcal{D}\int d^3\vecx \;W_{R}(\vecx)\delta (\vecx) \int d\hatr_1 \;W_{R}(\vecx+\vecr_1) \delta(\vecx+\vecr_1) \nonumber \\ 
&  \;\times \int d\hatr_2 \; W_{R}(\vecx+\vecr_2)\delta(\vecx+\vecr_2)\int d\hatn \; \PP_{111}(\hatn,\hatr_1,\hatr_2) \PP^*_{\ell_n\ell_1\ell_2}(\hatn, \hatr_1,\hatr_2) \nonumber \\ 
&  = \mathcal{D} \sum_{m_1,m_2=-\ell_1,-\ell_2}^{\ell_1,\ell_2}\sum_{m_n,m'_1,m'_2=-1}^1 \mathcal{C}_{m_n,m_1,m_2}^{1,\ell_1,\ell_2}\mathcal{C}_{-m_n,m'_1,m'_2}^{1,1,1}\nonumber \\ 
& \quad \times\int d^3\vecx \;\delta (\vecx) \;a_{\ell_1,m_1;1,m'_1;R}(\vecx,r_1)\;a_{\ell_2,m_2;1,m'_2;R}(\vecx,r_2)\;\dK_{\ell_n,1}\dK_{m_n,-m'_n}, 
\end{align} 
In the second equality, the $\hatn$ integral has been evaluated, yielding the Kronecker deltas that enforce the allowed angular structure. To perform the angular integrations, we have used the isotropic basis functions \cite{cahn_iso} 
\begin{align} \label{eq:Iso_function_def}
&\PP_{L_1L_2L_3}(\hatr_1,\hatr_2,\hatr_3) = \sum_{M_1,M_2,M_3} \mathcal{C}_{M_1,M_2,M_3}^{L_1,L_2,L_3} Y_{L_1M_1}(\hatr_1)Y_{L_2M_2}(\hatr_2)Y_{L_3M_3}(\hatr_3), 
\end{align} 
where $\mathcal{C}$ is defined as: 
\begin{align} 
\mathcal{C}_{m,m',m^"}^{\ell,\ell',\ell^"}\equiv(-1)^{\ell+\ell'+\ell^"} \tj{\ell}{\ell'}{\ell^"}{m}{m'}{m^"}. 
\end{align} 
We have also introduced
\begin{align}\label{eq:continous_a_lm} 
a_{L_i,M_i;1,M'_i;R}(\vecx, r_i) \equiv \int d\hatr_i \;W_{R}(\vecx+\vecr_i)\delta(\vecx+\vecr_i) \;Y^*_{L_iM_i}(\hatr_i) Y_{1M'_i}(\hatr_i), 
\end{align} 
which collects the angular dependence associated with each secondary leg of the triangle.

\section{Discrete estimator}\label{sec:Discrete_Estimator}
Real galaxy surveys involve discrete particle positions, so we now write the estimator in its discrete-particle form (we use the term galaxies and particles interchangeably here). Let the catalog contain galaxies at positions $\vecx_i$ with weights $w_i$. For a primary particle $i$, we define the separation vector to a secondary particle $j$ as 
\begin{align} 
\vecr_{ij}\equiv \vecx_j-\vecx_i, \qquad r_{ij}\equiv |\vecr_{ij}|, \qquad \hatr_{ij}\equiv \frac{\vecr_{ij}}{r_{ij}} . 
\end{align} 
To pass from the continuous estimator to a particle estimator, we replace the angular integral appearing in the continuous quantity $a_{\ell m;1m'}(\vecx,r)$ by a sum over secondary particles around each primary. Since the data are discrete, the radial separation between the primary galaxy and the secondary galaxies is binned into shells. For a radial bin $a$, with lower and upper edges $r_a^{\rm min}$ and $r_a^{\rm max}$, we define the binning function 
\begin{align} 
\Theta_a(r) \equiv 
\begin{cases} 
1, & r_a^{\rm min}\leq r < r_a^{\rm max},\\ 0, & \text{otherwise}. 
\end{cases} 
\end{align} 
The radially-binned discrete analog of $a_{\ell m;1m'}(\vecx,r)$, \cref{eq:continous_a_lm}, is then 
\begin{align}\label{eq:discrete_a_lm_odd3pcf} 
a^{a}_{\ell m;1m';R}(\vecx_i) \equiv \sum_{j\neq i}^{N_{\rm G}} \;W_{R}(\vecx_i+\vecr_{ij})\;w_j\, \Theta_a(r_{ij})\, Y^*_{\ell m}(\hatr_{ij})\, Y_{1m'}(\hatr_{ij}), 
\end{align} 
with $N_{\rm G}$ the number of galaxies. This quantity is the discrete version of the angular projection over the leg, restricted to radial shell $a$. We note that the spherical harmonic factor $Y_{1m'}(\hatr_{ij})$ arises from the parity-odd kernel proportional to $\PP_{111}(\hatn,\hatr_1,\hatr_2)$. 

Using \cref{eq:discrete_a_lm_odd3pcf}, the continuous parity-violating 3PCF estimator, \cref{eq:Odd3PCF_Continuous_Estimator}, becomes 
\begin{align}
\label{eq:discrete_odd_3pcf_estimator} 
&\widehat{\zeta}^{ab}_{1,\ell_1,\ell_2} = \mathcal{D}\, \sum_i^{N_{\rm G}} w_i \sum_{m_n,m_1,m_2} \sum_{m'_1,m'_2} \mathcal{C}_{m_n,m_1,m_2}^{1,\ell_1,\ell_2} \mathcal{C}_{-m_n,m'_1,m'_2}^{1,1,1}\nonumber \\
&\qquad \qquad \times a^{a}_{\ell_1m_1;1m'_1;R}(\vecx_i)\; a^{b}_{\ell_2m_2;1m'_2;R}(\vecx_i). 
\end{align} 
Around a given primary particle indexed by $i$, one first computes the single-secondary quantities in \cref{eq:discrete_a_lm_odd3pcf} for each radial shell, and then combines the corresponding shell coefficients through the angular-coupling factors. This gives a separable estimator: no explicit loop over pairs of secondary particles is required once the $a^{a}_{\ell m;1m';R}(\vecx_i)$ have been computed. If the radial bins $a$ and $b$ are distinct, \cref{eq:discrete_odd_3pcf_estimator} will not contain contributions from the same galaxy appearing on both secondary legs. As first demonstrated in \cite{se_3pt_alg}, this approach computes the 3PCF in $\mathcal{O}(N_{\rm G}^2)$ time.

\section{Correcting for Survey Geometry}
\label{sec:Edge_Corrections}
Correcting for survey geometry is a fundamental ingredient in cosmological clustering analyses because real galaxy surveys never sample the Universe uniformly or completely. When estimating NPCFs, galaxies near the survey edges have fewer available neighbors simply due to geometry, not physics, leading to incorrect measurements. Edge-correction schemes, \cite{kayo_2004, szapudi_szalay, landy_szalay, kerscher_2000}, ensure that measured correlations reflect intrinsic cosmic structure rather than survey geometry. For higher-order statistics such as the 3PCF and 4PCF, where configurations are more sensitive to complex boundary effects, accurate edge correction is especially critical, as even small mismodeling of the survey footprint can propagate into significant systematic errors \cite{encore}. Following \cite{szapudi_szalay,se_3pt_alg}, we compute the edge correction for the 3PCF model presented above as
\begin{align}
    \widehat{\zeta}(\hatn,\vecr_1,\vecr_2)
    =
    \frac{NNN(\hatn,\vecr_1,\vecr_2)}
    {RRR(\hatn,\vecr_1,\vecr_2)} ,
\end{align}
where $N\equiv D-\alpha R$ denotes the data-minus-random field, with $D$ and $R$ denoting the weighted data and random catalogs, respectively. The normalization factor $\alpha$ matches the total weighted number of randoms to the total weighted number of galaxies,
\begin{equation}
    \alpha \equiv
    \frac{\sum_{i\in D} w_i}
    {\sum_{j\in R} w_j},
\end{equation}
where the sums are taken over the survey sample or region being analyzed. The numerator and denominator are then expanded in the same isotropic basis used for the parity-violating 3PCF,
\begin{align}
    NNN(\hatn,\vecr_1,\vecr_2)
    =
    \sum_{L_n,L_1,L_2}
    \mathcal{N}_{L_n,L_1,L_2}(r_1,r_2)\;
    \PP_{L_n L_1 L_2}(\hatn,\hatr_1,\hatr_2),
\end{align}
and
\begin{align}
    RRR(\hatn,\vecr_1,\vecr_2)
    =
    \sum_{\lambda_n,\lambda_1,\lambda_2}
    \mathcal{R}_{\lambda_n,\lambda_1,\lambda_2}(r_1,r_2)\;
    \PP_{\lambda_n \lambda_1 \lambda_2}(\hatn,\hatr_1,\hatr_2).
\end{align}
Similarly, the edge-corrected 3PCF is expanded as
\begin{align}
    \widehat{\zeta}(\hatn,\vecr_1,\vecr_2) = \sum_{\ell_n,\ell_1,\ell_2}\widehat{\zeta}_{\ell_n,\ell_1,\ell_2}(r_1,r_2)\;\PP_{\ell_n \ell_1 \ell_2}(\hatn,\hatr_1,\hatr_2).
\end{align}
Using $NNN=\widehat{\zeta}\,RRR$, we obtain
\begin{align}
&\sum_{\ell_n,\ell_1,\ell_2}\sum_{\lambda_n,\lambda_1,\lambda_2}\widehat{\zeta}_{\ell_n,\ell_1,\ell_2}(r_1,r_2)\;\mathcal{R}_{\lambda_n,\lambda_1,\lambda_2}(r_1,r_2)\;\PP_{\ell_n\ell_1\ell_2}(\hatn,\hatr_1,\hatr_2)\;\PP_{\lambda_n\lambda_1\lambda_2}(\hatn,\hatr_1,\hatr_2) \nonumber \\
& \quad \;\;= \sum_{L_n,L_1,L_2}\mathcal{N}_{L_n,L_1,L_2}(r_1,r_2)\;\PP_{L_n L_1 L_2}(\hatn,\hatr_1,\hatr_2).
\end{align}
To project this equation onto a single basis element, we multiply by $\PP^*_{L_n L_1 L_2}(\hatn,\hatr_1,\hatr_2)$ and integrate over the angular variables. This gives
\begin{align}
    \mathcal{N}_{L_n,L_1,L_2}(r_1,r_2) = \sum_{\ell_n,\ell_1,\ell_2}\sum_{\lambda_n,\lambda_1,\lambda_2}\widehat{\zeta}_{\ell_n,\ell_1,\ell_2}(r_1,r_2)\;\mathcal{R}_{\lambda_n,\lambda_1,\lambda_2}(r_1,r_2)\;\mathcal{G}^{L_n,L_1,L_2}_{\ell_n,\ell_1,\ell_2;\lambda_n,\lambda_1,\lambda_2},
\end{align}
where the coupling coefficients are defined by 
\begin{align}
    &\mathcal{G}^{L_n,L_1,L_2}_{\ell_n,\ell_1,\ell_2;\lambda_n,\lambda_1,\lambda_2} \equiv \int d\hatn\,d\hatr_1\,d\hatr_2\;\PP_{\ell_n\ell_1\ell_2}(\hatn,\hatr_1,\hatr_2)\nonumber \\
    & \qquad \qquad \qquad \quad \;\times\PP_{\lambda_n\lambda_1\lambda_2}(\hatn,\hatr_1,\hatr_2)\;\PP^*_{L_n L_1 L_2}(\hatn,\hatr_1,\hatr_2).
\end{align}
The explicit form for this constant is given in Eq. (46) of \cite{cahn_iso} and Eq. (E.6) of \cite{Ortola_4PCF}. This expression describes how the survey geometry mixes different, $(\ell, \ell',\ell^")$, angular multiplets. Separating the monopole of the random counts, $\mathcal{R}_{0,0,0}$, we have
\begin{align}
    \frac{\mathcal{N}_{L_n,L_1,L_2}(r_1,r_2)}{\mathcal{R}_{0,0,0}(r_1,r_2)} = \sum_{\ell_n,\ell_1,\ell_2}\mathcal{M}^{L_n,L_1,L_2}_{\ell_n,\ell_1,\ell_2}(r_1,r_2)\;\widehat{\zeta}_{\ell_n,\ell_1,\ell_2}(r_1,r_2),
\end{align}
with the edge-correction matrix
\begin{align}
    \mathcal{M}^{L_n,L_1,L_2}_{\ell_n,\ell_1,\ell_2}(r_1,r_2) \equiv \mathcal{G}^{L_n,L_1,L_2}_{\ell_n,\ell_1,\ell_2;0,0,0} + \sum_{\substack{\lambda_n,\lambda_1,\lambda_2 \\ \lambda_n+\lambda_1+\lambda_2>0}}\frac{\mathcal{R}_{\lambda_n,\lambda_1,\lambda_2}(r_1,r_2)}{\mathcal{R}_{0,0,0}(r_1,r_2)}\;\mathcal{G}^{L_n,L_1,L_2}_{\ell_n,\ell_1,\ell_2;\lambda_n,\lambda_1,\lambda_2}.
\end{align}
In matrix notation, this becomes
\begin{align}
    \frac{\boldsymbol{\mathcal{N}}(r_1,r_2)}{\mathcal{R}_{0,0,0}(r_1,r_2)} = \boldsymbol{\mathcal{M}}(r_1,r_2)\;\widehat{\boldsymbol{\zeta}}(r_1,r_2),
\end{align}
where and therefore the edge-corrected multiplets are 
\begin{align}
    \widehat{\boldsymbol{\zeta}}(r_1,r_2) = \boldsymbol{{\mathcal{M}}}^{-1}(r_1,r_2)\;\frac{\boldsymbol{\mathcal{N}}(r_1,r_2)}{\mathcal{R}_{0,0,0}(r_1,r_2)}.
\end{align}
In the limit of a boundary-freee survey, all multiplets of the randoms save for the monopole vanish, and the correction matrix, $\boldsymbol{\mathcal{M}}$, reduces to the monopole contribution alone. For a realistic survey, the higher random multiplets encode both angular mask and radial selection function, and the matrix inversion removes the induced mixing between the measured multiplets.

\section{Discussion $\&$ Conclusions}
\label{sec:conclusion}

In this work, we have introduced a parity-odd 3PCF estimator designed to test for PV in the large-scale distribution of galaxies. The 4PCF is the lowest-order statistic capable of probing parity violation in a fully 3D isotropic field. However, when constrained to a 2D spherical shell, a triangle cannot be rotated into its mirror image and thus can probe PV. This makes it possible to construct a parity-odd 3PCF once the triangle is dotted with an externally-defined shell normal. 

We have also derived an edge-correction scheme appropriate for this parity-odd 3PCF. As in standard NPCF analyses, the survey geometry can mix angular modes and generate biased estimates if not properly corrected. By expanding both the data-minus-random triple counts and the random triple counts in the same angular basis, the effect of the survey mask can be written as a mode-coupling matrix. The edge-corrected 3PCF multiplets are then obtained by inverting this matrix for each radial-bin pair. 

A practical advantage of the parity-odd 3PCF is that it retains the separable structure of fast 3PCF estimators. Once the single-secondary angular coefficients have been accumulated around each primary galaxy, the estimator does not require an explicit loop over all pairs of secondary galaxies. For fixed angular truncation and radial binning, the resulting scaling is therefore $\mathcal{O}(N_G^2)$, as in the standard spherical-harmonic 3PCF algorithm of \cite{se_3pt_alg}. This scaling should not be interpreted as an advantage over the 4PCF measurements, since algorithms such as \textsc{encore} can compute the 4PCF in $\mathcal{O}(N_G^2)$ time \cite{encore}. Rather, the advantage is that the statistic is a 3-point quantity: it has fewer angular couplings, and a potentially smaller data vector than the parity-odd 4PCF. These features may make the estimator computationally cheaper in practice and easier to use as an independent consistency test.

A second advantage is that the parity-odd 3PCF may provide a useful cross-check of parity-odd 4PCF measurements with different covariance and systematic sensitivities. Existing 4PCF parity-violation searches have shown that the inferred detection significance is very sensitive to the covariance treatment and to the agreement between the data and the mocks (as discussed in \Cref{sec:intro}). These results motivate complementary parity-odd statistics whose covariance properties and systematic responses differ from those of the 4PCF. 

There are, however, a few considerations that are worth keeping in mind when applying this estimator to data. First, the definition of the shell normal is central to the construction. As discussed in the main text, the vector $\hatn$ is taken to be an externally-defined normal vector associated with the spherical shell. This choice is essential for ensuring that the resulting scalar probes fundamental PV. For example, if $\hatn$ were constructed based on the internal triangle geometry, such as $\hatr_1 \times \hatr_2$, \Cref{eq:pv_kernel} would yield a parity-even quantity. In this sense, the estimator is fully well-defined once the spherical shell and $\hatn$ are defined.

The first future goal is to validate the estimator on simulations. Using\texttt{Quijote} simulations, we will complete such a validation, since these simulations include both parity-odd and parity-even initial conditions. The procedure is to construct artificial spherical shells within the simulation box by choosing observer positions and selecting particles or halos in radial intervals around those observers. The parity-odd 3PCF can then be measured on these shells. A successful proof-of-concept would show that the estimator gives a null result for parity-even simulations, and finds a signal for parity-odd simulations. 

The second future goal is to apply this estimator to real galaxy survey data. Such an application will be validated alongside realistic mocks that include survey geometry, redshift-space distortions, observational weights, radial selection, and fiber assignment. We note that measuring both the parity-odd 3PCF and 4PCF and finding agreement between the two statistics would strengthen the case for a physical parity-odd signal, while disagreement could help identify survey systematics, or covariance-related issues in existing measurements.

The final and most important goal is to provide the estimator in a public code and assemble it into a fully reproducible analysis pipeline. This implementation should be informed by the validation steps above, ensuring that all required components---shell construction, angular coefficient accumulation, radial binning, and random-catalog edge correction---are integrated in a consistent and efficient framework. 

Overall, this paper should be viewed as a first step toward a new test of parity violation in large-scale structure. The estimator developed here provides the analytical foundation, and it can be applied to current and future galxy surveys such as the LSST, DESI, DESI-Extension, Roman and SPHEREX.

\acknowledgments
WOL's work is supported by the National Science Foundation Graduate Research Fellowship under Grant No.\ DGE-2236414. Any opinions, findings, and conclusions or recommendations expressed in this material are those of the author and do not necessarily reflect the views of the National Science Foundation. ZS acknowledges funding from NASA grant number 80NSSC24M0021 and funding from UF Research AI award \#00133699.

\appendix

\bibliographystyle{JHEP}
\bibliography{references}

\end{document}